\documentclass[usenatbib]{mnras}

\input epsf

\usepackage{graphicx}
\usepackage{txfonts}
\usepackage{epsfig}

\title[Sensitivity of low-degree solar p modes]{Sensitivity of low-degree solar p modes to active and ephemeral regions: frequency shifts back to the Maunder Minimum}

\author[W.J. Chaplin et al.]{William J. Chaplin$^{1,2}$, Rachel Howe$^{1,2}$,
  Sarbani Basu$^{3}$, Yvonne Elsworth$^{1,2}$, \newauthor 
  Timothy W. Milbourne$^{4,5}$, Rapha\"elle D. Haywood$^{5,6}$,  Guy R. Davies$^{1,2}$, \newauthor Steven J.~Hale$^{1,2}$, Andrea Miglio$^{1,2}$, Eddie Ross$^{1,2}$\\ $^{1}$School of Physics and Astronomy,
  University of Birmingham, Birmingham, B15 2TT, United
  Kingdom\\ $^{2}$Stellar Astrophysics Centre (SAC), Department of
  Physics and Astronomy,Aarhus University,\\ Ny Munkegade 120, DK-8000
  Aarhus C, Denmark\\ $^3$Department of Astronomy, Yale University, PO
  Box 208101, New Haven, CT, 065208101, USA\\ $^{4}$Department of Physics, Harvard University, 17 Oxford Street, Cambridge MA 02138, USA\\ $^{5}$Harvard-Smithsonian Center for Astrophysics, Cambridge, MA 02138, USA\\ $^{6}$NASA Sagan Fellow}

\begin{document}

\maketitle

\begin{abstract}

We explore the sensitivity of the frequencies of low-degree solar p-modes to near-surface magnetic flux on different spatial scales and strengths, specifically to active regions with strong magnetic fields and ephemeral regions with weak magnetic fields. We also use model reconstructions from the literature to calculate average  frequency offsets back to the end of the Maunder minimum. We find that the p-mode frequencies are at least three times less sensitive (at 95\,\% confidence) to the ephemeral-region field than they are to the active-region field. Frequency shifts between activity cycle minima and maxima are controlled predominantly by the change of active region flux. Frequency shifts at cycle minima (with respect to a magnetically quiet Sun) are determined largely by the ephemeral flux, and are estimated to have been $0.1\,\rm \mu Hz$ or less over the last few minima. We conclude that at epochs of cycle minimum, frequency shifts due to near-surface magnetic activity are negligible compared to the offsets between observed and model frequencies that arise from inaccurate modelling of the near-surface layers (the so-called surface term). The implication is that this will be the case for other Sun-like stars with similar activity, which has implications for asteroseismic modelling of stars.

\end{abstract}




\section{Introduction}
\label{sec:intro}

It has long been established that solar p-mode frequencies respond to the Sun's changing levels of near-surface magnetic activity (e.g., \citealt{woodard85, palle89, elsworth90, libbrecht90}), and hence provide a probe of the underlying physical changes driving these variations. An aspect that has been little explored is the relative response of the frequencies to small-scale, weak magnetic fields and large-scale, strong field. Here, we use helioseismic data from the Birmingham Solar-Oscillations Network (BiSON) \citep{hale16} on globally coherent, low angular degree (low-$l$) modes, together with spatially resolved magnetogram data, to analyse the sensitivity of the modes to strong-field, active-region (AR) and weak-field, ephemeral-region (ER) flux. By disentangling the response to each component, we are able for the first time to make predictions of the extent to which observed frequencies are shifted with respect to a magnetically quiet or grand-minimum like state by near-surface field. We also use model reconstructions from the literature of the photospheric magnetic flux to calculate average frequency offsets dating back to the end of the Maunder minimum. 

There is relevance here for asteroseismic inference on the fundamental properties of other Sun-like stars. The default approach for such analyses fails to account for activity-dependent changes to the frequencies over time, and/or that frequencies are offset from the field-free scenario computed by standard stellar evolutionary models.  Having a better handle on the sensitivity of p modes to strong and weak-component field is also relevant for understanding the seismic variability shown by Sun-like stars (e.g., see \citealt{garcia10, salabert16, kiefer17, santos18}), where available data are likely to sample targets showing a different mix of strong and and weak activity components (e.g., see also \citealt{montet17, raddick18}).

\section{Background}
\label{sec:back} 

Extended, strong magnetic dipoles form active regions (AR), and give rise to sunspots and plage. The number of, and flux contained within, AR varies strongly across the solar cycle. Smaller, weaker magnetic dipoles form ephemeral regions (ER) -- containing so-called network field -- which also show a relationship to the emergence of AR \citep{harvey00}. Although there is some disagreement in the literature over whether this weaker component varies with the cycle \citep[see e.g.,][and references therein]{vieira10}, some studies (e.g. see \citealt{harvey94}) show a clear positive correlation, which may be assumed to be driven in part by the decay and dispersal of magnetic field associated with AR. Some of the photospheric magnetic flux is also dragged out into the heliosphere, giving rise to open flux, which has contributions from both AR and ER \citep{krivova07}.

The p-mode frequency shifts are driven by some combination of direct and indirect effects due to the magnetic field (e.g., \citealt{gough90,dziembowski05}). The field can act directly, through the Lorentz force, which scales with the strength of the field; or indirectly, by changing the stratification, which scales with magnetic pressure, i.e., the square of the field. The geometry of the field also matters. If the filling factor of the field in the near-surface layers is low, then the mean absolute field is proportional to the mean-square field \citep{woodard91}, meaning the indirect effect would have the same dependence on the field as the direct effect. It is perhaps then not surprising that analyses of observed shifts have failed to discriminate between a linear or quadratic dependence on the field. Here, we make the reasonable assumption that use of a linear dependence is valid and that total magnetic fluxes in the AR and ER components determine the frequency shifts shown by the globally coherent low-degree modes. (Note a quadratic dependence on the field would inevitably reduce the implied contribution to the shifts from the weak-field ER component: our adopted linear model may be regarded as providing an upper limit to the ER sensitivity.)

\section{Sensitivity analysis}
\label{sec:anal}

Estimates of the surface magnetic flux in AR and ER were extracted from magnetogram data collected at the Wilcox Solar Observatory (WSO; \citealt{duvall77}), following the basic approach outlined in \citet{arge02}. We considered the data in re-mapped 5-degree-square patches. We divided into strong (AR) and weak (ER) components using the criterion that the patches had magnetic field strengths above or below a given threshold in units of Gauss. The total flux estimates then correspond to the area-weighted total unsigned longitudinal field strength above or below the threshold (as appropriate) per Carrington rotation, given by summing the Carrington-rotation averages over latitude. 


\begin{figure*}
	\centering
	\includegraphics[width=0.6\textwidth]{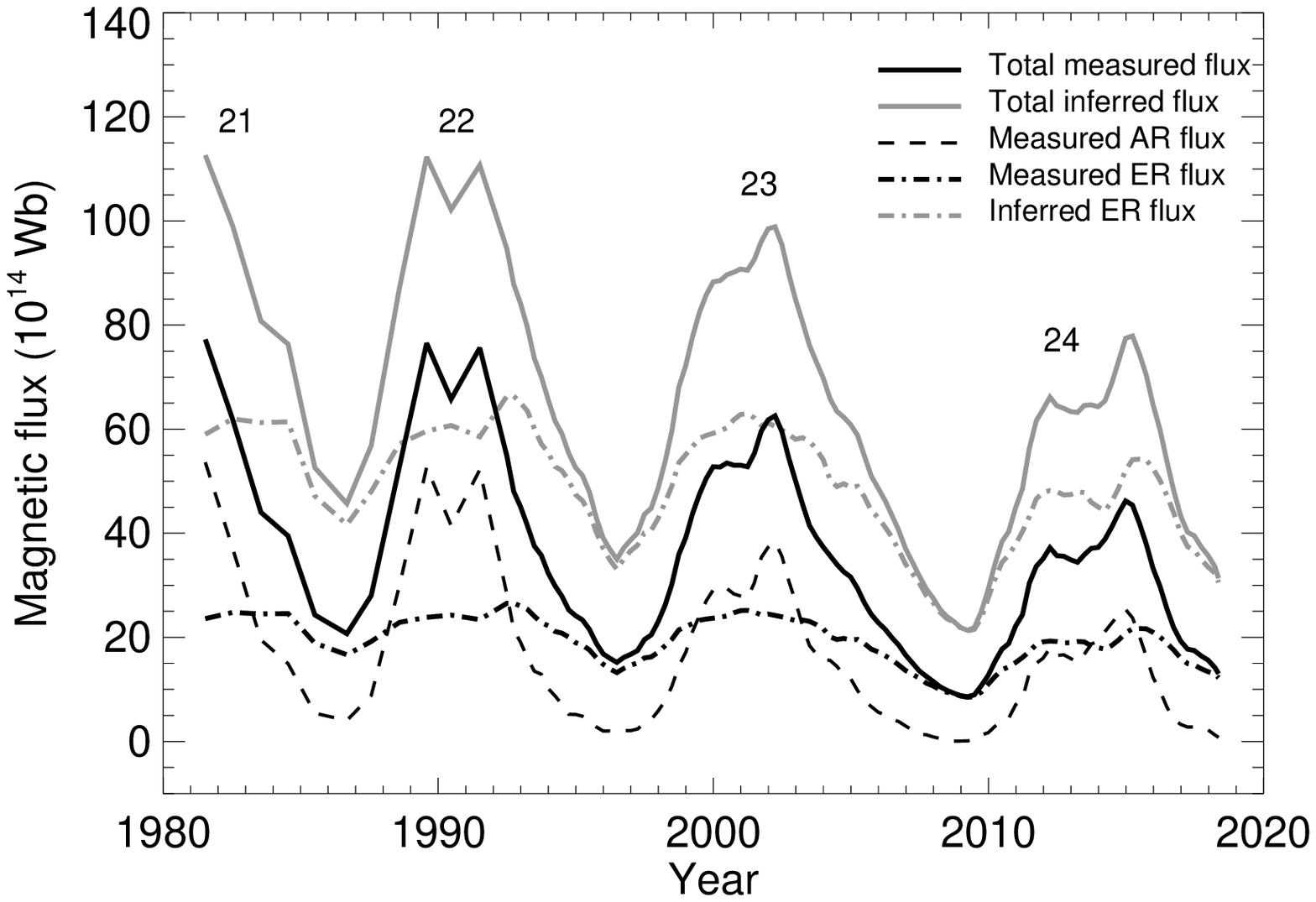}
	\includegraphics[width=0.6\textwidth]{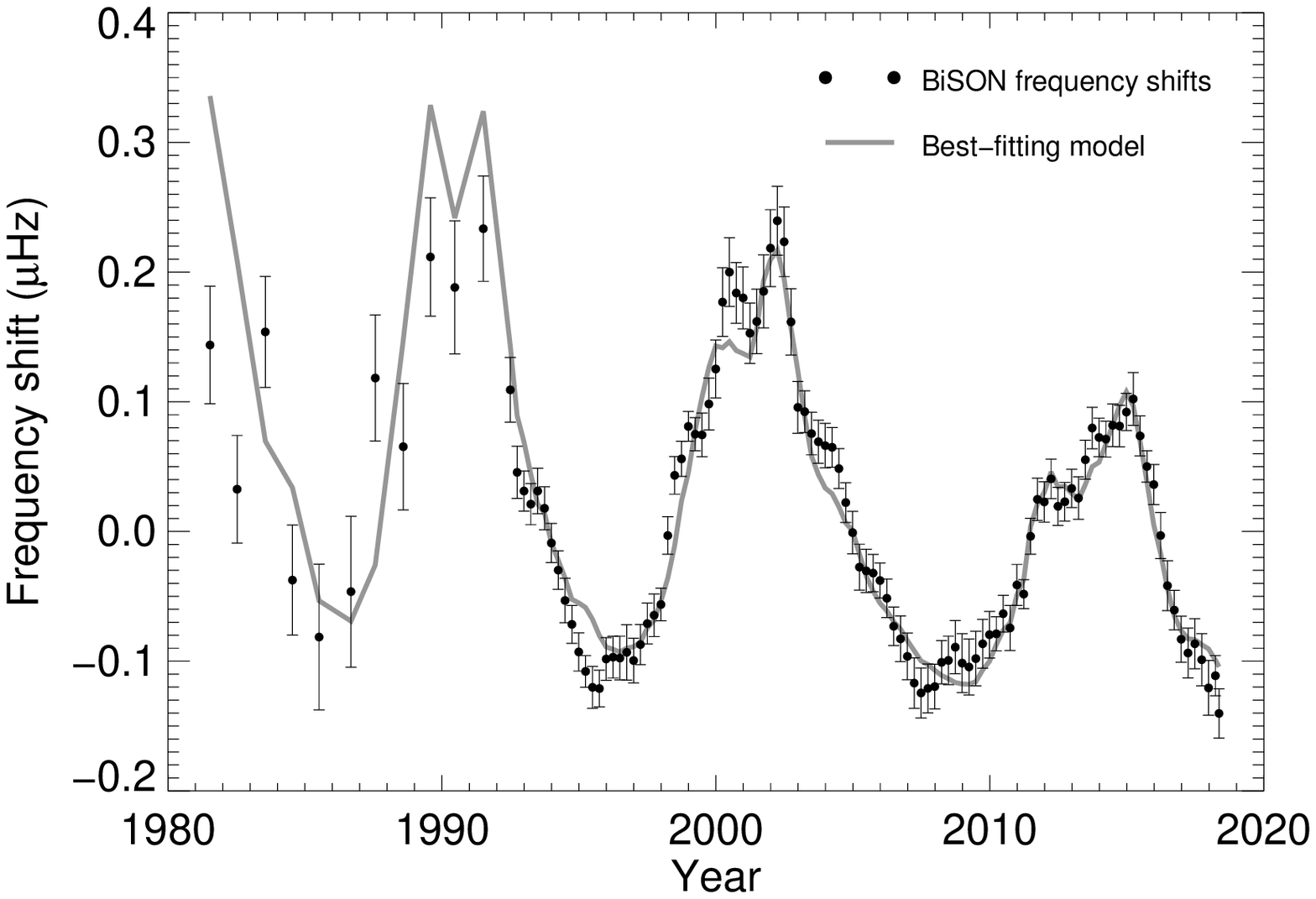}
	\caption{Top panel: Measured AR (black dashed line) and ER (black dot-dashed line) fluxes, and total measured flux (black solid line). The grey dot-dashed line shows the inferred total ER flux, i.e., having taken into account that not all the ER flux is captured by the observations. The inferred total flux (the sum of the measured AR and inferred ER) is shown in solid grey. Also shown are the numbers of each activity cycle. Bottom panel: BiSON mean frequency shifts (points with error bars) and modelled frequency shifts (solid line) given by using the best-fitting coefficients to Equation~\ref{eq:dnu}.}
	\label{fig:flux1}
\end{figure*}


We tested different thresholds from 10 to 30\,G and settled on a division at 15\,G, which gave results on the ER and AR fluxes in good agreement with those of \citet{harvey94} and \citet{tapping07}. Note that our threshold is lower than the 25\,G threshold adopted by \citet{harvey94}, likely owing to the different intrinsic spatial resolutions of the respective datasets and differences in the procedures used. We must take into account that not all the ER flux is captured by the observations, owing to their finite resolution \citep{krivova04}. Here, we have assumed that the observations captured a fraction $\beta = 0.4$ of the actual ER flux (see, e.g., \citealt{krivova07, vieira10}). 

The top panel of Fig~\ref{fig:flux1} shows the  measured AR flux, $F_{\rm AR}(t)$ (black dashed line) and  measured ER flux, $F_{\rm ER}(t)$ (black dot-dashed line). The total measured flux (sum of the measured AR and ER fluxes) is shown by the solid black line. The grey dot-dashed line shows the inferred total ER flux, after applying the $\beta$ correction. The inferred total flux, shown by the solid grey line, then corresponds to $F_{\rm AR}(t) + \beta^{-1}\,F_{\rm ER}(t)$. As we shall see below, our main conclusions are fairly insensitive to the choice of $\beta$ (though we take uncertainties in the parameter into account).

In order to provide an additional check on our inferred fluxes, we also analysed higher-resolution magnetograms obtained by the Helioseismic and Magnetic Imager (HMI) onboard the Solar Dynamics Observatory (SDO; \citealt{Schou2012}, \citealt{Pesnell2012}). We used thresholding algorithms developed by \citet{Fligge2000} and later used for solar radial-velocity modelling by, for example, \citet{Meunier2010} and \citet{Haywood2016}; further details may be found in \citet{Milbourne2019} and Haywood et al (in prep). Note the HMI data span only cycle 24. This independent analysis gave AR and ER variations in very good agreement with contemporaneous WSO results.

Our goal is to determine the sensitivity of low-degree (low-$l$) solar p-mode frequencies to the AR and ER components. We used mean frequency shifts calculated from data collected by BiSON spanning the second half of Schwabe cycle 21, all of cycles 22 and 23, and most of cycle 24, i.e., a period of over 37\,years. The mean shifts -- from 1-year timeseries offset by 3\,months -- correspond to averages taken over 28 modes spanning degrees $l=0$ to 2 and frequencies from 2292 to $3703\,\rm \mu Hz$. Further details may be found in \citet{howe17}. We fitted these mean shifts $\delta\nu(t)$ to the linear model:
 \begin{equation}
 \delta\nu(t) = c_0 + c_1 \left[ F_{\rm AR}(t) + \alpha \left(\frac{F_{\rm ER}(t)}{\beta}\right)\, \right],
 \label{eq:dnu}
 \end{equation}
where $c_0$ and $c_1$ correspond to the zero- and first-order model coefficients, and $\alpha$ reflects the relative sensitivity of the mode frequencies to the AR and ER flux. Fits were made by performing a multiple linear regression on the two independent variables $F_{\rm AR}(t)$ and $F_{\rm ER}(t)$ to find the best-fitting $c_0$, $c_1$ and $\alpha$. The bottom panel of Fig.~\ref{fig:flux1} shows the mean frequency shifts (points with error bars) and modelled frequency shifts (solid line) given by using the best-fitting coefficients. Note the seismic data coverage is sparser prior to full roll-out of BiSON in 1992 as a six-site network.

Before we come to discuss the results from the model fit, it is already apparent from visual inspection of Fig.~\ref{fig:flux1} that the frequencies must be significantly less sensitive to the ER component than they are to the AR component. The total flux at the cycle 23/24 minimum is noticeably lower than at the cycle 22/23 minimum, and most of this drop comes from a significant reduction in the ER flux. Yet the average frequencies show hardly any change (see also \citealt{broomhall17}). Contrast this with the differences shown between the adjacent cycle minima and the cycle 23 maximum. The ER and AR fluxes change by similar amounts -- each by roughly $30 \times 10^{14}\,\rm Wb$ -- and we see a $\approx 0.3\,\rm \mu Hz$ change in frequency. Were the frequencies to be equally sensitive to the ER and AR components, we would then expect a frequency change of around $\approx 0.08\,\rm \mu Hz$ between the two minima. However, the change we see is negligible. 

Our linear regression analysis gave $\alpha = 0.11 \pm 0.09$ (stat) $\pm 0.02$ (sys), the fit indeed confirming that the p-mode frequencies are significantly less sensitive to the ER component than they are to the AR component, i.e., at least three times less sensitive at 95\,\% confidence.  The systematic uncertainty on $\alpha$ takes into account an uncertainty on the sensitivity coefficient $\beta$ of $\simeq 0.1$. The best-fitting model also gave a sensitivity coefficient of $c_1 = 7.9 \pm 0.7$ (stat)\,nHz per $10^{14}\,\rm Wb$. We tested the robustness of the above inference by demonstrating we could recover accurate parameters from artificial datasets made to mimic the BiSON and magnetogram data, having underlying values of $\alpha$ ranging from zero to unity.

\citet{santos16} estimated that only 30\,\% of the observed p-mode frequency shifts can be attributed to sunspots. Whilst this might seem at odds with our result, a direct comparison is difficult since the other component (contributing 70\,\% of the shifts) that Santos et al. considered contains contributions from both AR and ER flux (sunspots of course contributing to the former).

\section{Discussion}
\label{sec:disc}

Next, we consider the implications of the above results for the p-mode frequency shifts we would expect with respect to a magnetically quiet Sun or a Maunder-minimum (grand-minimum) like state.

Our results show that frequency shifts between activity cycle minima and maxima are controlled predominantly by the change of AR flux. However, frequency shifts of modes at cycle minima -- by which we mean frequency offsets with respect to a magnetically quiet Sun -- are determined largely by the ER component, since the AR flux can then be all but absent. The best-fitting coefficients of Equation~\ref{eq:dnu} imply that at epochs corresponding to the last few cycle minima, the average frequency offset was likely to have been around or possibly below $0.1\,\rm \mu Hz$. This is at a level of one third or less of the total frequency change shown by the modes.  We might reasonably conclude that the seismic behaviour of the Sun at cycle minimum will therefore approximate that shown by a magnetically quiet star (bearing also in mind that the magnetic flux will be distributed spatially in a fairly homogeneous way). The offset is certainly much smaller than the offsets between observed frequencies and predictions from solar evolutionary models that arise from inaccurate modelling of the near-surface layers, i.e., the so-called ``surface term'' (e.g., see \citealt{jcd86, kjeldsen08, ball14}). There remains the possibility that more deeply buried magnetic field could also change the frequencies \citep{gough90b, kiefer18} by amounts comparable to the surface term offset.


\begin{figure*}
	\centering
	\includegraphics[width=0.6\textwidth]{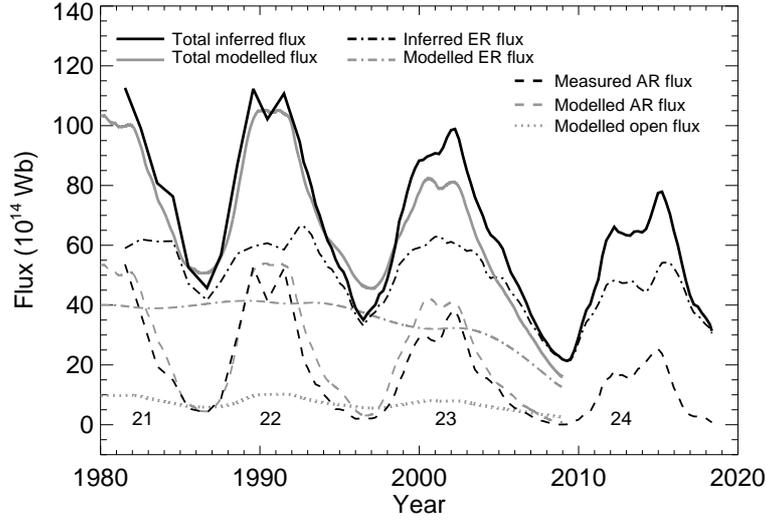}
	\caption{Measured/inferred fluxes (from the magnetograms) [in black] and model predictions of the fluxes (from \citealt{vieira10}) [in grey]. The AR fluxes are plotted with dashed lines, the ER fluxes with dot-dashed lines, and the total fluxes with solid lines. The model prediction includes an explicit contribution from open flux, shown here by the dotted grey line.}
	\label{fig:flux3}
\end{figure*}


We have also used model predictions by \citet{vieira10} (see also \citealt{solanki02, krivova07}) of the Sun's magnetic flux to calculate an average p-mode frequency shift, due to the near-surface fields, dating back to the end of the Maunder minimum. These estimates of course rest on the fidelity of the model predictions. Fig.~\ref{fig:flux3} compares our measured and inferred fluxes with the model predictions of \citet{vieira10}. That there is good overall agreement is not surprising, since the total flux predicted by the model was tested against observed total fluxes. 

We see reasonable agreement between our measured AR flux (shown by the black dashed line) and the model flux (shown by the grey dashed line). Our inferred ER flux is plotted as the grey dot-dashed line; recall this is our measured ER flux corrected explicitly for the impact of the finite spatial resolution of the magnetograms (with $\beta=0.4$). The model ER is shown in grey (same line style). The inferred ER shows a more pronounced variation with the solar cycle, and a higher overall mean level (by about 25\,\%). That said, the model ER is seen to provide a reasonable match to the lower envelope of our inferred ER.  One possible reason for the mismatch concerns assumptions made regarding $\beta$ (see previous discussions). But another reason is the contribution from open flux. As noted in Section~\ref{sec:back} above, it is assumed that open flux will be captured by the magnetogram data in both the ER and AR components. Given the much better match of the measured and model AR fluxes, it is possible that a sizeable fraction of the open flux is captured by our measured ER flux (thereby explaining the positive offset here with respect to the modelled ER level). Note the model prediction of the open flux is plotted in Fig.~\ref{fig:flux3} with a dotted grey line. Our inferred total flux is shown by the black solid line, with the model predicted total flux -- given by the sum of the model AR, ER and open fluxes -- in grey.


\begin{figure*}

 \centerline {\includegraphics[width=0.5\textwidth]{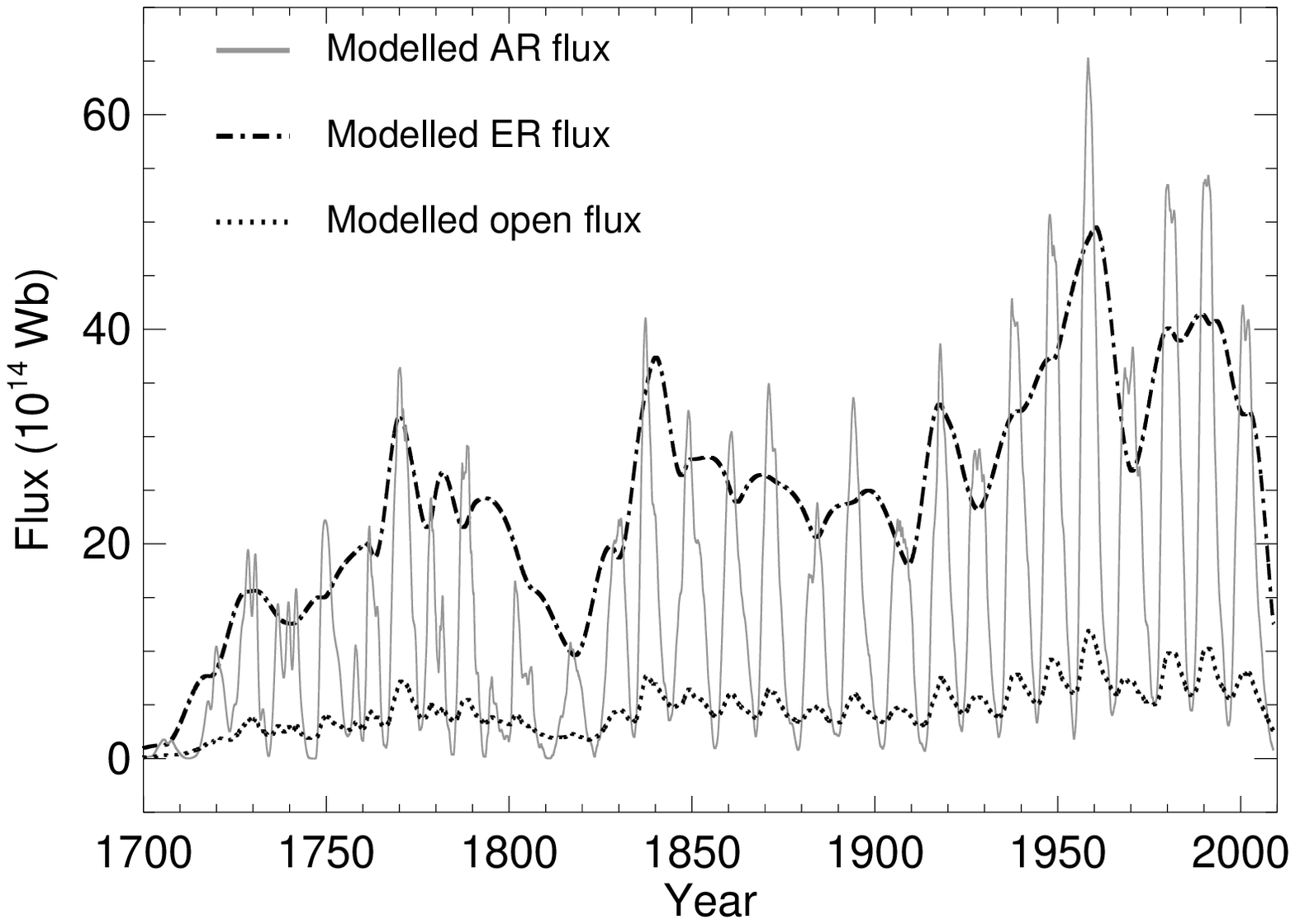}
              \includegraphics[width=0.5\textwidth]{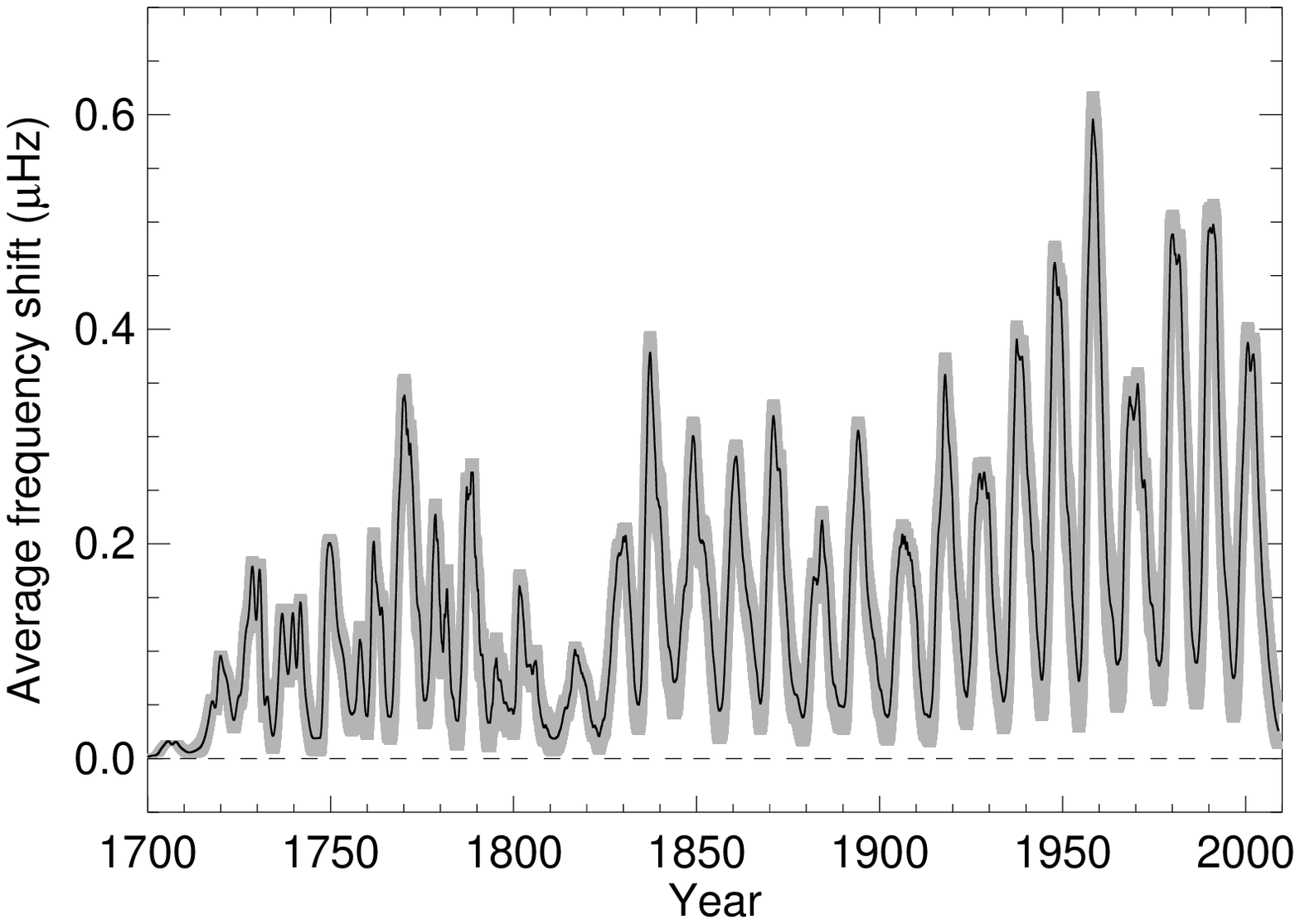}}
	\caption{Left-panel: \citet{vieira10} prediction of the AR, ER and open flux over the last 300 years. Right-hand panel: predicted absolute frequency shift implied by use of Equation~\ref{eq:dnu}, with the ER and AR fluxes in the top panel used as inputs with the coefficients fixed to the above-mentioned best-fitting values. The error envelope captures uncertainties on the best-fitting model parameters, and uncertainty over how the open flux is captured by the magnetograms (see text).} 

	\label{fig:tot1}
\end{figure*}


The left-hand panel of Fig.~\ref{fig:tot1} shows the \citet{vieira10} prediction of the AR, ER and open flux over the last 300 years. The right-hand panel is our estimate of the average p-mode frequency shift over this extended period, as implied by use of Equation~\ref{eq:dnu} with the fluxes in the left-hand panel used as inputs and with the coefficients fixed to the aforementioned best-fitting values. The grey error envelope captures uncertainties on the best-fitting model parameters, and uncertainty over the open flux contribution. With regards to the latter, we calculated frequency shifts for two extreme cases: one where we assumed the open flux acts like ER flux, so that the p modes are assumed to show the same sensitivity to both components; and another where instead we assumed the open flux acts like AR flux.

We see calculated frequency offsets at modern cycle minima of around $0.1\,\rm \mu Hz$. Moving back in time, the offsets diminish. Swings in frequency between cycle minima and maxima also reduce in size. Note at the earliest epochs, which are at the end of the Maunder minimum, even the ER flux -- and hence the calculated shifts -- tends to zero. This is a result of the model predictions being conditioned on sunspot number to estimate various flux contributions: in sum, there is no model AR flux present to seed the model ER flux, and any latent model ER flux decays away on a much shorter timescale than the length of the Maunder minimum.


\section*{Acknowledgements}

We would like to thank all those who are, or have been, associated with BiSON, in particular P. Pall\'e and T. Roca-Cortes in Tenerife and E. Rhodes Jr. and S. Pinkerton at Mt. Wilson. BiSON is funded by the Science and Technology Facilities Council (STFC), under grant ST/M00077X/1. W.J.C., G.R.D., Y.E. and E.R. acknowledge the support of the Science and Technology Facilities Council (STFC). S.B. acknowledges support from NSF grant AST-1514676. R.H. acknowledges support from the University of Birmingham. Funding for the Stellar Astrophysics Centre is provided by The Danish National Research Foundation (Grant agreement no.:DNRF106). This work was performed under contract with the California Institute of Technology (Caltech)/Jet Propulsion Laboratory (JPL) funded by NASA through the Sagan Fellowship Program executed by the NASA Exoplanet Science Institute (R.D.H.). A.M. acknowledges the European Research Council (ERC) under the European Union's Horizon 2020 research and innovation programme (project ASTEROCHRONOMETRY, grant agreement no. 772293). Wilcox Solar Observatory data used in this study was obtained via the web site \url{http://wso.stanford.edu}.  The Wilcox Solar Observatory is currently supported by NASA. The HMI data used are courtesy of NASA/SDO and the HMI science team, and are publicly available at \url{http://jsoc.stanford.edu/}. SDO is part of the Living with a Star Program within NASA's Heliophysics Division. The authors thank Sami Solanki for providing data from \citet{vieira10}.


\end{document}